\documentclass[12pt]{article}
\usepackage{amsfonts,amsmath,amssymb,amsthm,bbold}
\usepackage{scalerel}[2016/12/29]
\usepackage{epsfig,graphicx}
\usepackage{tikz}
\usepackage{float}
\usepackage{braket}
\usepackage{url,hyperref}
\usetikzlibrary{decorations,arrows}
\usetikzlibrary{decorations.markings}
\usetikzlibrary{shapes.geometric}

\newcommand{\bb}{\hskip -0.1cm}

\def\bb{\hskip -0.5mm}
\def\be{\begin{equation}}
\def\ee{\end{equation}}
\def\bea{\begin{eqnarray}}
\def\eea{\end{eqnarray}}

\def\rme{\mathrm{e}}

\def\rme{\mathrm{e}}

\begin{document}

\title{\bf{ Inclusion statistics and particle condensation in 2 dimensions}}

\date{\today}

\author{St\'ephane Ouvry$^{*}$  \and Alexios P. Polychronakos$^{\dagger}$}

\maketitle

\begin{abstract}
We propose a new type of quantum statistics, which we call inclusion statistics,  in which particles tend to coalesce
more than ordinary bosons. Inclusion statistics is defined in analogy with exclusion statistics, in which statistical 
exclusion is stronger than in Fermi statistics, but now extrapolating beyond Bose statistics, resulting in
statistical inclusion. A consequence of inclusion statistics is that the lowest space dimension in which 
particles can condense in the absence of potentials is $d=2$, unlike $d=3$ for the usual Bose-Einstein 
condensation. This reduction in the dimension happens for any inclusion stronger than bosons,
and the critical temperature increases with stronger inclusion. Possible physical realizations of inclusion statistics involving attractive interactions
between bosons may be experimentally achievable.
\end{abstract}

\noindent
* LPTMS, CNRS,   Universit\'e Paris-Saclay, 91405 Orsay Cedex, France\\
${ }$\hskip 0.38cm{\it stephane.ouvry@u-psud.fr}

\noindent
$\dagger$ Physics Department, the City College of New York, New York, NY 10031, and\\
${ }$\hskip 0.35cm The Graduate Center of CUNY, New York, NY 10016, USA
\\ \noindent
${ }$\hskip 0.38cm{\it apolychronakos@ccny.cuny.edu}

%\tableofcontents
\section{Introduction}

Bose-Einstein condensation is a highly nontrivial manifestation of quantum statistics: bosons condense in 3 dimensions when they are cooled below a critical temperature, whereas fermions never condense due to Pauli exclusion. A spectacular experimental confirmation \cite{exp,ket} of this stunning theoretical prediction has been achieved in 1995, and has since been reproduced by many groups. A salient feature of Bose-Einstein condensation is that,
in the absence of external potentials, it is possible only in dimension $d\ge 3$.
 
On the other hand, and on another front, it is known \cite{LM} that in  $2$ dimensions intermediate statistics between
Bose and Fermi statistics are possible due to the nontrivial  topology of the configuration space of identical particles, i.e., to the topologically nontrivial braiding of spacetime paths of particles going around each other.
It follows that the
quantum $N$-body wavefunction acquires a multivalued phase,
which, when  gauged away, results in particles with ordinary Bose statistics but interacting { through
a two-body} Aharonov-Bohm-type term, meaning that particles are endowed with a fictitious charge $e$ and a
fictitious flux {$\phi$}. The statistical parameter is $g ={\phi/\phi_o}$ where the flux quantum is
$\phi_o=h/e$. This is the famous anyon model that interpolates from
 % singular gauge transformation  since it does not preserve the magnetic field 
Bose statistics ($g=0$) to Fermi statistics ($g=1$), and, by periodicity of the phase, returns to Bose statistics
when $g=2$.    

A  major hurdle in the anyon model  is that  the $N$-body problem is not solvable except for a class of exact eigenstates, the so-called linear states, which manifest themselves when a constant magnetic field and/or a harmonic 
potential are added. 
To make progress, a simplification is needed:
one can add a  magnetic field perpendicular to the plane, couple the anyons to it, and
project the system on its lowest Landau level (LLL). The  LLL-anyon model  happens to be
solvable, since there is a complete set of LLL-anyon states (the linear states)
that interpolate between the LLL-Bose and LLL-Fermi basis states. Then, from the  LLL-anyon spectrum, one
can readily get the LLL-anyon thermodynamics \cite{Das}, and as a consequence, the
LLL-anyon occupation number, that is, the number of anyons per available one-body state, which in this degenerate case is nothing but the filling factor in the LLL. 

Furthermore, aside from the microscopic anyonic formulation and space dimensionality considerations, an interesting feature  of the LLL-anyon thermodynamics is that  it allows for a  combinatorial reformulation
in terms of the occupation of single-particle quantum states, which leads to  the so-called Haldane exclusion
statistics \cite{Haldaneha}. Indeed, the  LLL-anyons thermodynamics  turn out to be identical to those
derived from Haldane statistics \cite{wu}, defined purely in terms of Hilbert space counting arguments.
This combinatorial reformulation extends to positive values of the
exclusion parameter $g$ beyond the periodicity range $g\in[0,2[$,  {i.e., $g\ge 2$}. This results to an interparticle exclusion stronger than the
standard Pauli exclusion, justifying the term exclusion statistics. 
 
In this note we propose,  instead of going beyond Fermi statistics, to go beyond Bose statistics by
considering negative values of the exclusion parameter, { i.e., $g< 0$}. 
To do this, we take the combinatorial reformulation at face value and analytically continue it to negative
values of $g$.  This leads to particles for which the degeneracy of states {\it increases} when they
occupy neighboring states, and therefore have a propensity to include, rather than exclude each other.
We thus call this extension of quantum statistics {\it inclusion} statistics.

In the sequel we will examine the properties of particles obeying inclusion statistics and
derive their
thermodynamics. We will see that an interesting physical consequence of these statistics is a
lowering  of the lowest dimension at which particles can condense {compared to the Bose case}
from $d=3$ to $d=2$. We leave to a separate publication \cite{sep} the technical details pertaining to inclusion statistics, and in particular the case of a discrete  one-body spectrum.

\section{$g$-exclusion thermodynamics \label{noway}}

\noindent
Following Haldane's Hilbert space counting argument\footnote{ As stated in the introduction the $g$-exclusion thermodynamics can be as well directly obtained from the microscopic LLL-anyon model.},
exclusion statistics of order $g>0$ { can be}  formulated by postulating that the degeneracy $G$ of $N$ particles
occupying $K$ degenerate states is
{\be G_g (K,N)={[K-(g-1)(N-1)]!\over N![K-g (N-1)-1]!}\nonumber\ee}
Clearly $g=0$ reproduces the bosonic result ${K+N-1 \choose N}$ and $g=1$ the fermionic one
${K \choose N}$. For integer $g\ge 0$ this can be interpreted as particles placed on a linear
one-body spectrum with the constraint that no more than 1 particle can occupy any set of $g$
adjacent states. In this sense, each particle ``excludes" other particles from occupying its nearby states.
An alternative definion of the degeneracy can be adopted \cite{Poly}
\be
G_g (K,N) = {K [K-(g-1) N-1]! \over N! (K-g N)!}
\label{cir}\ee
which corresponds to placing particles in a periodic spectrum, i.e., on a circle. The two definitions are equivalent in the
$K \gg 1$ limit and lead to the same thermodynamics, but the second one is more convenient for
deriving this limit.
Note that for integer $g > 0$ there is a finite number of particles $\lfloor K/g\rfloor$ that can be
placed in $K$ states, beyond which $G_g(K,N)$ vanishes, which implies the upper bound for the filling
fraction $N/K \le 1/g$.

The grand partition function for exclusion-$g$ particles in $K$ states of energy $\epsilon$, at inverse
temperature $\beta$ and chemical potential  $\mu$
\be
{\cal Z} (K,z) = \sum_{N=0}^\infty G_g(K,N) z^N ~,~~~ z = e^{\beta (\mu - \epsilon)}\nonumber
\ee
becomes extensive in the thermodynamic limit $K \gg 1$ and acquires the form
\be
\ln {\cal Z} = K \ln y + {\cal O}(K^{-1})\nonumber
\ee
(in fact, for the circular counting (\ref{cir}) the perturbative corrections ${\cal O}(K^{-1})$ vanish and there
are only nonperturbative corrections of order ${\cal O}(e^{-K})$). The function $y$ can, thus, be interpreted as
an effective grand partition function for $g$-exclusion particles at chemical potential $\mu$ on a single
state with energy $\epsilon$. It satisfies the equation
\be
y^g-y^{g-1}=z \label{estate}\ee
The physically relevant solution of this equation is for $y>1$. 
The single-state cluster coefficients $c_n$, defined by
\be
\ln y = \sum_{n=1}^\infty c_n z^n\nonumber
\ee
are derived to be
\be c_{n}={1\over n!} \prod_{k=1}^{n-1} (k-n g) = {(-1)^{n-1}\over ng}{n g \choose n}
\nonumber\ee
the second expression valid for $g\ge 1$. The $c_n$ never vanish for integer $g\ge 0$ and alternate in sign
for $g\ge 1$.

All this generalizes to the grand partition function $Z$ for particles with $g$-exclusion and 1-body
density of states $\rho(\epsilon)$
\be
\ln Z=\int_0^{\infty}\rho(\epsilon)\ln y\;d\epsilon
\nonumber \ee
where we assumed without loss of generality that $\rho(\epsilon)$  starts at $\epsilon=0$.
The occupation number at energy $\epsilon$ follows from
\be n = {1\over \beta} {\partial \over \partial\mu} \ln y =
z{ \partial\over \partial z} \ln y\nonumber
\ee
which, in combination with (\ref{estate}), implies
\be
n = {y-1 \over g y + 1 -g} ~,~~~ z={n\over (1+(1-g)n)^{1-g}(1-gn)^{g}}
\label{n}\ee
This gives the grand potential as
\be\nonumber\ln Z=\beta P V =\int_0^{\infty}\rho(\epsilon)\ln \bigg(1+{n\over 1-g n}\bigg)\;d\epsilon
\label{part} \ee
which determines the equation of state and reaffirms that there is a maximum critical occupation
number
\be n\le   1/g
\nonumber\ee
For bosons ($g=0$) there is no maximum, while for fermions ($g=1$) we have the standard Pauli
exclusion principle with at most one fermion per quantum state. For $g>1$ we have that the critical 
occupation number should be smaller than $1$, corresponding to exclusion.

Other thermodynamic quantities, such as the particle number $N$, energy $E$, and entropy $S$
are given by
\be\begin{split}
&N=\int_0^{\infty}\rho(\epsilon)\; n \;d\epsilon ~,~~~~~E=\int_0^{\infty}\rho(\epsilon)\; n\;\epsilon\;d\epsilon \\
&S= \ln Z+\beta E- \beta \mu N = \int_0^{\infty}
\rho(\epsilon)\ln{(1+n(1-g))^{1+n(1-g)}\over (1-n g)^{1-n g}\;\;n^{n}}d\epsilon
\end{split}\nonumber\ee

\section{Inclusion statistics and particle condensation \label{previous}}
 
As we stated in the introduction, we define inclusion statistics by taking $g<0$ in the
expressions of section (\ref{noway}). The combinatorial formulae for the number of states,
expression for cluster coefficients, and equation for the single-level grand partition function remain
the same, understanding that $g$ is negative.

Noting that the equation for $y$ can be rewritten as
\be
\left({1\over y}\right)^{1-g} - \left({1\over y}\right)^{(1-g)-1} = -z
\nonumber\ee
we deduce the relation between $g$-statistics and $(1-g)$-statistics
\be
y(z,1-g)={1\over y(-z,g)}\label{pivot}\ee
Therefore, inclusion statistics of order $g<0$ are related to exclusion statistics of order $1-g>1$
upon changing the sign of the grand potential and of the fugacity. In particular,
the cluster coefficients $c_n$ remain the same but now become all positive:
\be\nonumber
c_n (g) = (-1)^{n-1} c_n (1-g) >0 ~~~\text{for}~~~ g<0
\ee

The crucial new feature of the $g<0$ case is that the branch of the equation corresponding to
$z>0$ admits real solutions only for $z$ below some maximal value, $z<z_{max}$. Indeed,
the left-hand-side
$y^g - y^{g-1} = z$ of (\ref{estate}), viewed as a function of $y$, vanishes for $y=1$ (as it should)
and has a critical point at $y=y_c $ with
\be\nonumber
y_c = {g-1\over g}
\ee
For $g<0$ this occurs at a physical value of $y >1$ and corresponds to a maximum, at which  $z$  takes its maximum value $z_{max}$  
\be\nonumber
z_{max} = {(-g)^{-g} \over (1-g)^{1-g}} >0
\ee
Therefore, for $g<0$ the solution $y=1+z+\cdots >1$ exists as long as $0<z < z_{max}$. For
each such $z$, (\ref{estate}) has two solutions, but only the smaller one, $y<y_c$, reached
from $y=1$ and $z=0$ by increasing $z$, is physical.

This is already enough to demonstrate that a system of inclusion particles will undergo
condensation at a nonzero temperature. Near $z_{max}$
\be
z- z_{max} \simeq - C (y-y_c)^2 ~~~\Rightarrow~~~ y \simeq y_c - \sqrt{z_{max}-z \over C}
\nonumber\ee
with
\be\nonumber
C = {(-g)^{3-g} \over 2 (1-g)^{2-g}} >0
\ee
and the average particle number $n=z{\partial\over \partial z} \ln y$ 
\be n \simeq {z_{max} \over 2 y_c \sqrt{C(z_{max} - z)}}
\nonumber\ee
%This will be relevant for Bose condensation.
%The total number of particles $N$ in a macroscopic system is
%\be
%N=\int_0^\infty n(\epsilon) \, \rho(\epsilon) d\epsilon
%\nonumber\ee
The maximal number of particles that the system can accommodate under
a normal thermodynamic distribution (without condensation) 
is achieved for $\mu =\mu_{max}$ at $\epsilon = 0$, that is
\be
z_{max} = e^{\beta \mu_{max}} ~~~\Rightarrow ~~~
\mu_{max} = {1\over \beta} \ln z_{max}
\nonumber\ee
So, if
\be\label{Nmax}
N_{max} := \int_0^\infty n(\epsilon)|_{\mu=\mu_{max}} \, \rho(\epsilon) d\epsilon < \infty
\ee
the system will undergo particle condensation when  the actual number of particles $N > N_{max}$.
It follows that condensation will occur if the  integral (\ref{Nmax}) does not diverge. The integral cannot diverge as
$\epsilon \to \infty$ since $n(\epsilon)|_{\mu=\mu_{max}}$ goes to
zero exponentially in that limit, so it can only diverge as $\epsilon \to 0$, that is, $z \to z_{max}$.
For $\mu = \mu_{max}$, $z = z_{max} e^{-\beta \epsilon}$ and
\be\nonumber
n \simeq {{\sqrt {z_{max}}}\over 2y_c \sqrt{C (1-e^{-\beta \epsilon}) }} 
\ee
and thus, for $\epsilon \to 0$
\be\nonumber
n \sim \epsilon^{-1/2}
\ee
This has to be contrasted to the behavior of $n$ for bosons ($g=0$) which is
$n \sim \epsilon^{-1}$.
The weaker dependence $n \sim \epsilon^{-1/2}$ sets in as soon as $g<0$
and implies a change in the critical dimension in which condensation happens.
Assuming the standard density of states 
\be
\rho(\epsilon)=V{c_d (2m)^{d/2}\over 2 h^d}\epsilon^{d/2-1}
\nonumber\ee
for free nonrelativistic particles in a volume $V$ in $d$ dimensions, with $c_d$ is the 
$d$-dimensional spherical factor $c_d={2 \pi^{d/2}/ \Gamma(d/2)}$,
%i.e., $\rho(\epsilon) \sim \epsilon^{d/2 - 1}$,
 we see that the integrand in (\ref{Nmax}) for $N_{max}$ near $\epsilon=0$ behaves as $\sim \epsilon^{(d-3)/2}$
and the integral will be finite as long as
\be {d-3 \over 2} > -1 ~~~\Rightarrow ~~~ d>1\nonumber
\ee
So we will have particle condensation  at $d=2$, i.e., one dimension lower than for 
standard Bose-Einstein condensation.

\section{Critical temperature for $g$-inclusion particle condensation}

To set the stage, let us first review the usual {\bf $g=0$} Bose-Einstein condensation
where  $\mu_{max}=0~\Rightarrow~ z_{max}=1$. The occupation number for $\mu = \mu_{max} = 0$ is
\be %z\rme^{-\beta\epsilon}={n\over (1-gn)^{g}(1+(1-g)n)^{1-g}}\Rightarrow \rme^{-\beta\epsilon}={n\over 1+n}\Rightarrow 
n={1\over \rme^{\beta\epsilon}-1}
\nonumber\ee
so
\bea 
\hskip -1cm N_{max} =\bb
\int_{0}^{\infty}\bb\bb n(\epsilon)\rho(\epsilon)d\epsilon = V{c_d (2m)^{d/2}\over 2 h^d}\bb
\int_{0}^{\infty}\bb {\epsilon^{d/2-1}\over \rme^{\beta\epsilon}-1}d\epsilon &=& \infty \quad\quad
\quad\quad\text{ if }d\le 2 \nonumber\\ 
&=&{V\over \lambda^d}\zeta (d/2)\quad\, \text{ if }d>2
\nonumber\eea
where we introduced the thermal wavelength $\lambda={h\over \sqrt{2\pi m kT}}$ and used
\bea
\int_0^{\infty}{u^{d/2-1}\over \rme^u-1}\, du\nonumber &=&\Gamma(d/2)\text{Li}_{d/2}(1)=\Gamma(d/2)\zeta (d/2)
\eea
%and introduced the thermal wavelength $\lambda={h\over \sqrt{2\pi m kT}}$.
The critical temperature $T_c$ (for $d>2$) is defined as the temperature at which $N=N_{max}$, that is,
\be
N={V\over \lambda_c^d}\zeta(d/2)
\nonumber \ee
which gives
\be T_c= 
{h^2\over 2\pi m k}  \bigg({\rho\over \zeta(d/2) }\bigg)^{2/d}
\label{critical}\ee
where $\rho = N/V$ is the boson density. Clearly, if
\be
T\le T_c ~~~\Rightarrow~~~ N_{max} = {V\over \lambda^d}\zeta (d/2)\le N
\nonumber\ee
So $N_{cond}=N-N_{max}$ particles will condense in the ground state and there is a macroscopic
Bose-Einstein condensate of a fraction of particles $N_{cond}/N$
\be
{N_{cond}\over N} =  1- {V\over N \lambda^d}\zeta (d/2)
= 1-\bigg({ T \over T_c}\bigg)^{d/2}
\nonumber\ee

Let us now turn to 
 $g$-inclusion ($g<0$). At the maximal fugacity $z_{max}={(-g)^{-g} (1-g)^{g-1}}$   (\ref{n}) gives
\be
\rme^{\beta\epsilon}={(-{1\over g}+n)^{g}({1\over 1-g}+n)^{1-g}\over n}
=\left({1\over ng}-1\right)^{g} \left({1\over n(1-g)}+1\right)^{1-g}
\nonumber\ee
Performing a change of integration variable in the integral 
$N=\int_{0}^{\infty} n(\epsilon)\rho(\epsilon)d\epsilon$ from $\epsilon$ to $t=1/n$
\bea\nonumber
%\beta \epsilon &=& g\ln\left(n-{1\over g}\right) +(1-g)\ln\left(n+{1\over 1-g}\right) -\ln n \cr
\beta \epsilon &=&  g\ln\left(1-{t\over g}\right) +(1-g)\ln\left(1+{t\over 1-g}\right)
\eea
\be \beta d\epsilon = {t \over (t-g)(t+1-g)} dt
\nonumber\ee
we obtain
%\bea
% \int_{0}^{\infty} n(\epsilon)\rho(\epsilon)d\epsilon &=& V{c_d (2m)^{d/2}\over 2 h^d} \beta^{-d/2} \int_0^\infty dn
%{\left[ g\ln\left(1-{1\over ng}\right) +(1-g)\ln\left(1+{1\over n(1-g)}\right)\right]^{d/2 -1} \over 
%\left(1-gn\right)\left(1+(1-g)n\right)}\nonumber\\&=& {V\over \lambda^d}\int_0^\infty dn
% {1\over\Gamma(d/2)}{\left[ g\ln\left(1-{1\over ng}\right) +(1-g)\ln\left(1+{1\over n(1-g)}\right)\right]^{d/2 -1} \over 
%\left(1-gn\right)\left(1+(1-g)n\right)}\label{toto}
%\eea
\be\nonumber
N_{max} =  {V\over \lambda^d}{\tilde{\zeta}}(g,d/2)
\ee
where we defined
\be\label{clear}
{\tilde{\zeta}}(g,s) = {1\over\Gamma(s)} \int_0^\infty dt\, {\left[ g\ln\left(1-{t\over g}\right) 
+(1-g)\ln\left(1+{t\over 1-g}\right)\right]^{s -1} \over 
(t-g)(t+1-g)}
\ee
For $g<0$, the integral  ${\tilde{\zeta}}(g,d/2)$ is always finite as $t \to \infty$, but the integrand behaves as
$\sim (t^2)^{d/2-1} = t^{d-2}$ as $t\to 0$, so it will be finite as long as $d>1$,
as found in  section (\ref{previous}).
%\be\int_{0}^{\infty} n(\epsilon)\rho(\epsilon)d\epsilon={V\over\lambda^d}{\tilde{\zeta}}(g,d)\nonumber\ee
From the above, the critical temperature for particle condensation in $d>1$ is derived as
\be
T_c={h^2\over 2\pi m k} \bigg({\rho\over {\tilde{\zeta}}(g,d/2)}\bigg)^{2/d}
\label{criticalbis}\ee
 Note that for $g>0$, for which there is no condensation, the integrand in
${\tilde{\zeta}}(g,d/2)$ develops poles within the integration domain.

The function ${\tilde{\zeta}}(g,s)$ defined in (\ref{clear}) (not to be confused with the Hurwitz zeta function
$\zeta(s,a)$) is a generalization of Riemann's zeta function, reducing  for $g=0$ to
the standard zeta function $\zeta(s)$
\be
{\tilde{\zeta}}(0,s) = {1\over\Gamma(s)}\int_0^\infty dt 
{\left[ \ln (t+1) \right]^{s -1} \over 
t(t+1)}= \zeta (s)
\nonumber\ee
This also recovers the bosonic limit $g=0$ as
\be\nonumber
 \int_{0}^{\infty} n(\epsilon)\rho(\epsilon)d\epsilon ={ V\over \lambda^d}
\zeta(0,d/2)= {V\over \lambda^d}\zeta (d/2)\quad \text{ if }d>2
\ee
${\tilde{\zeta}}(g,d/2)$ can actually be explicitly evaluated for
even dimensions $d=2,4,\dots$ in terms of polylogarithms. Specifically,
\bea\nonumber
{\tilde{\zeta}}(g,2/2) &=& -g\text{Li}_1\big({1\over g}\big)+(1-g)\text{Li}_1\big({1\over 1-g}\big)\cr
&=& \text{Li}_1\big({1\over 1-g}\big)=\ln \big(1-{1 \over g}\big)\cr
{\tilde{\zeta}}(g,4/2) &=& -g\text{Li}_2\big({1\over g}\big)+(1-g)\text{Li}_2\big({1\over 1-g}\big) \cr
&=&\text{Li}_2\big({1\over 1-g}\big)+{1\over 2}g\ln^2\big({1-{1\over g}}\big)\cr
{\tilde{\zeta}}(g,6/2) &=& -g\text{Li}_3\big({1\over g}\big)+(1-g)\text{Li}_3\big({1\over 1-g}\big)+{1\over 6}g(g-1)\ln^3\big({1-{1\over g}}\big)\cr &=&-g\bigg(\text{Li}_3\big({1\over g}\big)-(g-1)\text{S}_{1,2}\big({1\over g}\big)\bigg)-({g\to 1-g})\cr
{\tilde{\zeta}}(g,8/2) &=& -g\bigg(\text{Li}_4\big({1\over g}\big)+(g-1)\bigg({1\over 2}\text{Li}_2\big({1\over g}\big)^2-\text{S}_{2,2}\big({1\over g}\big)-(g-1)\text{S}_{1,3}\big({1\over g}\big)\bigg)\bigg)-({g\to 1-g})\cr
{\tilde{\zeta}}(g,10/2) &=& g^2 \bigg((2 g - 3) \text{Li}_5\big({1\over g}\big) +(g-1)\bigg( \text{Li}_{4}  \big({1\over g}\big)  \text{Li}_{1}  \big({1\over g}\big) - \text{S}_{3,2}\big({1\over g}\big) + (g - 1) \text{S}_{1,4} 
     \big({1\over g}\big)\bigg)\bigg)-({g\to 1-g})\nonumber
\eea
where $\text{Li}_{n}$ and  $\text{S}_{n_1,n_2}$  stand for the polylogarithm and generalized Nielsen polylogarithm functions respectively. %{and $-{g\to 1-g}$ means subtracting the same term with $g\to 1-g$}.
Note that ${\tilde{\zeta}}(g,2/2)=\ln(1-1/ g)$  indicates that in the Bose limit
$g\to 0^{-}$ the integral diverges logarithmically, recovering the fact that Bose-Einstein condensation
does not happen in two dimensions only marginally. What happens is that the critical temperature $T_c$
increases as $g$ approaches $0$ from below and diverges at $g=0$.

We conclude by giving an alternative expression for ${\tilde{\zeta}}(g,s)$ among the several rewritings
of the original integral (\ref{clear}), with a somewhat simpler integrand,
  changing variable $t=(g-1) g u /(1 + g u)$  
 \be 
 {\tilde{\zeta}}(g,s)={1\over\Gamma(s)}\int_0^{-\frac{1}{g}} \frac{\left[g \ln(1+ u)-\ln (1+g u)\right]^{s-1}}{1+u} \, du\label{amazing}
\ee
This expression
%\footnote{{\blue One can check  numerically that in  the Bose limit $g\to 0^{-}$  both 
%integrals (\ref{amazing}, \ref{amazingbis}) converge   to $\zeta(d/2)$ as it should (in both cases the 
%integrand is vanishing, with in the first case a domain of integration becoming infinite $u\in[0,\infty]$ and in 
%the second case a denominator  diverging at the upper limit of integration $v\in[0,-1]$).}} 
 has the advantage of being well-defined for all 
values of $g$ and thus provides an
analytic continuation of ${\tilde{\zeta}}(g,d/2)$ to $g>0$. Note also that the change of variables $v=-u/(1+u)$ leads
to 
\be 
{\tilde{\zeta}}(g,s)=-{1\over\Gamma(s)}\int_0^{-\frac{1}{1-g}} \frac{\left[(1-g) \ln(1+ v)-\ln (1+(1-g) v)\right]^{s-1}}{1+v} \, dv
\label{amazingbis}\ee
demonstrating that ${\tilde{\zeta}}(g,s)$ satisfies the relation
\be {\tilde{\zeta}}(g,s)=-{\tilde{\zeta}}(1-g,s)
\nonumber\ee
This can be explicitly verified in the expressions for ${\tilde{\zeta}}(g,2/2)$ up to ${\tilde{\zeta}}(g,10/2)$. This is rooted
at the relation (\ref{pivot}) between the grand partition functions $y(z,g)$ and $y(-z,1-g)$, and relates
the physical sector $y>1$ of $g<0$ inclusion  statistics to the unphysical\footnote{{ The analytical 
continuation above has to be understood in this way. For actual  $g>0$ exclusion statistics, as discussed in 
section (\ref{noway}), no particle condensation occurs at any $d$. }} sector $y<1$ of $g>0$. As such,
it is not a physically relevant connection, although it may have possible physical implications in the right
context.

\section{Conclusions}

The possibility to observe particle condensation in dimensions lower than 3 is inherently interesting and,
if an appropriate realization of inclusion statistics is achieved, experimentally testable. Even in 3 dimensions,
inclusion statistics would have observable physical consequences, as it would raise the critical temperature,
other parameters remaining the same\footnote{ In the sodium experiment \cite{ket} the gas is made of   $5\times10^5$ atoms at densities  $10^{14}$ per cm$^{3}$ for a critical temperature of $2\mu K$.}. One can compare the critical temperature  (\ref{criticalbis}) for ($g<0$) inclusion
condensation to the critical temperature  (\ref{critical}) for the usual ($g=0$) Bose condensation  for a same species of atoms at the same density. Denoting by $r(g)$  the ratio %$T_g / T_0$ 
of these two temperatures, one gets
\bea  r(-5)&=&8.12894\nonumber\\  r(-4)&=& 7.10599 \nonumber\\  r(-3)&=&6.00133 \nonumber\\ r(-2)&=& 4.77881
\nonumber\\ r(-1)&=& 3.35538\nonumber \eea
This renders the observation of condensation, which is quite challenging and nontrivial for bosons,
substantially easier. For example, already for $g=-1$ inclusion, we obtain a more than threefold increase
in  the critical condensation temperature compared to that for the usual Bose-Einstein condensation.

The appearance in  the critical temperature for inclusion statistics of a generalization
of the Riemann zeta function is intriguing and may have some interesting implications (the standard
zeta function has actually appeared in a phenomenological description of the FQHE \cite{Leclair}).
Also interesting is the appearance of unphysical yet exact symmetries in this thermodynamics.
The relation (\ref{pivot}) between $g$ and $1-g$, mapping physical to unphysical sectors of inclusion
and exclusion statistics, has already been noted. An additional relation is the duality
\be
{1\over y(z,g)}+{1\over y(z^{-1/g}, g^{-1} )} = 1
\nonumber\ee
which maps statistics of the same kind (inclusion or exclusion) but inverts $g$ and rescales either the
temperature or the energy. For exclusion statistics ($g>0$) this mapping inverts the
energy scale and constitutes a generalized particle-hole duality (note that for $g=1$ it becomes
a self-duality between ordinary fermions and holes). For $g<0$, on the other hand, it does not
invert the spectrum and can be interpreted as a temperature rescaling. However, it maps the
physical ($y<y_c$) sector of one statistics to the unphysical ($y>y_c$) sector of the other,
as can be seen, e.g., by talking $z=0$, in which case $y(0,g)=1$ maps to $y(0,g^{-1}) =\infty$.
So it is also unphysical in this respect. However, physical and ``unphysical" sectors of a system quite
often map to alternative versions of the system that are not perturbatively related, and the possibility
remains that such a connection is also at play with inclusion statistics.

The most important and physically relevant question remains the realization of inclusion statistics in
physical systems. It is often the case that nonstandard statistics is a manifestation of interactions
between particles with ordinary statistics. The Calogero model is the canonical example, in which
particles exhibit properties consistent with free particles obeying generalized statistics \cite{NRBos}.
It is plausible
that a similar description involving {\it attractive} two-body potentials would give rise to inclusion
particles. The condensation and other thermodynamic properties of such systems could in principle
be derived independently, but the statistical interpretation would be a more compelling and generic
approach, capturing the essential features of a class of such systems. Interacting cold atoms, the
workhouse of low-temperature condensation physics, may offer the most promising possibility.

\noindent {\bf Acknowledgements :}
%Comparing  the critical temperature in $d=3$  for  usual Bose  gazes to the predicted temperature for such  $d=2$ putative systems with  say $g=-1$ statistics.
\noindent A.P. acknowledges the support of a PALM grant and the hospitality of LPTMS, CNRS at Université Paris-Saclay (Faculté des Sciences d’Orsay), where this work was initiated. The work of A.P. was supported in part by NSF under grant NSF-PHY-2112729 and by PSC-CUNY under grants 65109-0053 and 6D136-0003.

\end{document}